\documentclass[12pt]{iopart}

\newtheorem{proposition}{Proposition}

\usepackage{iopams}
\usepackage{graphicx,amssymb}
\usepackage{color}

\usepackage{ifthen}

\usepackage[latin1]{inputenc}
\usepackage{amsfonts}

\begin{document}

\title{Painlev\'e-Gullstrand synchronizations in spherical symmetry}

\author{Alicia Herrero$^1$ and Juan Antonio Morales-Lladosa$^2$}

\address{$^1$\ Institut de Matem\`atica Multidisciplinar and Departament de
Matem\`atica Aplicada, Universitat Polit\`ecnica de Val\`encia, \\
E-46220 Val\`encia, Spain.}

\address{$^2$\ Departament d'Astronomia
i Astrof\'{\i}sica, Universitat de Val\`encia, \\E-46100 Burjassot,
Val\`encia, Spain.}

\ead{aherrero@mat.upv.es; antonio.morales@uv.es}

\begin{abstract}
A Painlev\'e-Gullstrand synchronization is a slicing of the
space-time by a family of flat spacelike 3-surfaces. For
spherically symmetric space-times, we show that a
Painlev\'e-Gullstrand synchronization only exists in the region
where $(dr)^2 \leq 1$, $r$ being the curvature radius of the isometry group
orbits ($2$-spheres). This condition says that
the Misner-Sharp gravitational energy of these 2-spheres is
not negative and has an intrinsic meaning in terms of the norm of
the mean extrinsic curvature vector. It also provides an algebraic
inequality involving the Weyl curvature scalar and the Ricci
eigenvalues. We prove that the energy and momentum densities associated with the
Weinberg complex of a Painlev\'e-Gullstrand slice vanish in these
curvature coordinates, and we give a new interpretation of these slices
by using semi-metric Newtonian connections. It is also
outlined that, by solving the vacuum Einstein's equations in a
coordinate system adapted to a Painlev\'e-Gullstrand
synchronization, the Schwarzschild solution is directly obtained
in a whole coordinate domain that includes the horizon and both
its interior and exterior regions.
\end{abstract}



\section{Introduction}
\label{intro}

It is known that Painlev\'e \cite{Painleve}, Gullstrand
\cite{Gullstrand} and (some years later)  Lema{\^{\i}}tre
\cite{Elmestre} used a non-orthogonal curvature coordinate system
which allows to extend the Schwarzschild solution inside its horizon, see Eq.
(\ref{PG-form}) below. In this coordinate system, from now on
called a Painlev\'e-Gullstrand (PG) coordinate system, the metric
is not diagonal, but asymptotically flat, and regular across
the horizon, and then,  everywhere non-singular up to $r=0$.
Furthermore, one has a very simple spatial
3-geometry: the space-time appears foliated by a synchronization of
flat instants (hereafter called PG synchronizations).%
\footnote{We use here the abbreviations `PG coordinates' and `PG synchronization'
by sake of simplicity, without any intention of misplacing the relevant contribution
by Lema{\^{\i}}tre \cite{Elmestre}, who clarified the coordinate character of the
`Schwarzschild singularity' obtaining an extended metric form for the Schwarzschild solution.
In fact, the main motivation in papers \cite{Painleve,Gullstrand,Elmestre} is rather different.
For historical remarks about this subject, and some physical interpretations
see \cite{MarPoisson,HamLis}.}
\label{fn:PGL}

Nowadays, there is an increasing interest in the study of this type
of synchronizations. For instance: (i) in connection with
astrophysical applications, by taking into account that the dynamics
of the gravitational collapse should be pursued beyond the
Schwarzschild radius in a PG coordinate system
\cite{Eardley,PapaFont,AdlerBCL,ZiKu}, (ii) in spherically symmetric
space-times (SSSTs), as a convenient initial condition preserved
under time evolution \cite{GuMur,HuQaSi}, (iii) in relativistic
hydrodynamics, when an effective Lorentzian metric is introduced
\cite{Lichnero}, (iv) in non-relativistic situations admiting a
Lorentzian description, namely, in `analog gravity models' (see, for
example, \cite{Unruh,Visser1,Bilic,BarceloLSV,BarceloLV,Perez}), and also
(v) in modeling the black hole geometry and its associated physics,
or to describe some quantum
effects by starting from a Hamiltonian formulation
\cite{KraWil,PaWil,Schutzhold,HuWin}. More physical issues about
the use of PG coordinates and their interpretation can be found
in \cite{MarPoisson,HamLis,Visser2,NielsenVisser}. For a description
of the causal character of the geometric elements (coordinate
lines, coordinate 2-surfaces and coordinate 3-surfaces) associated
with PG coordinates see \cite{ere05,ere07}.

The existence of PG synchronizations in SSSTs has been studied in
the static case considering that the induced metric has vanishing
Ricci tensor \cite{BeigSi}, and some specific constructions are
presented in \cite{QadSidd}. In more general cases, this existence
is usually taken for granted but, recently, a limitation to this ansatz
has been pointed out \cite{LinSoo2009}. However, as far as the authors
are aware, a definitive interpretation of this limitation as well
as the analysis of the domains where a PG synchronization exists
have not been done up to now.
Then, some related questions arise:
does every SSST admits a region of physical interest where a
synchronization by flat instants exists? and, what are the advantages
of adapting coordinates to a flat spatial 3-geometry?
The main contribution
of this paper is to state the above limitation clearly,
in a form that it is coordinate independent, and to provide its physical interpretation.

Generalized (but, in general, non-flat) PG synchronizations have
been constructed in Schwarzschild \cite{Lake},
Reissner-Nordstr$\ddot{\rm o}$m \cite{LinSoo2009}, and Kerr
\cite{Doran,Natario} geometries, and also in non-vacuum SSSTs, where
new insights in the study of gravitational collapse scenarios are
achieved (by evolving an initial 3-geometry \cite{LaskyLun,
LaskyLunBu}). However, here we restrict ourselves to flat
synchronizations in order to discuss their existence in SSSTs.

The paper is structured as follows.  Section \ref{sec2} is devoted
to introduce some general formulae for the induced geometry on
space-like hypersurfaces and surfaces in SSSTs. In Sec. \ref{sec3} the condition for
the existence of a PG synchronization in SSSTs is analyzed and
physically interpreted. In Sec. \ref{sec4} we write the components
of the Weinberg pseudotensor \cite{Wein} with respect to a PG
synchronization, and we prove that the energy and momentum
densities of each PG slice vanish. In Sec. \ref{sec5} we consider
the semi-metric connection (see \cite{Bel}) associated with a
spherically symmetric metric expressed in PG coordinates and
provide new insights on the Newtonian interpretation
of the properties exhibited by the Schwarzschild field in these
coordinates. Sec. \ref{sec6} deals with the $3+1$ decomposition of
the Einstein equations with respect to a PG synchronization.
By integration of the vacuum equations one recovers the extended Schwarzschild
metric in PG coordinates, including the region inside the
horizon. Finally, in Sec. \ref{sec7} we discuss the role that our results can play for
a better understanding of the geometry and physics in SSSTs. Some preliminary results
of this work were presented at the Spanish Relativity meeting ERE-2009 \cite{ere09}.

Let us precise the used notation. The curvature
tensor $R^k _{\ lij}$ of a symmetric connection $\nabla$ is defined according with
the identity $\nabla_i \nabla_j \xi^k - \nabla_j \nabla_i \xi^k = R^k _{\ lij} \xi^l$
for the vector field $\xi$, and $R_{ij} \equiv R^k _{\ ikj}$
is the Ricci tensor. We take natural units in which $c=G=1$ and the Einstein
constant is $\kappa=8\pi$. We say that $\{t, r, \theta,
\varphi\}$ is a curvature coordinate system if
for constant $t$ and $r$ the line element is
$dl^2 = r^2 d\Omega^2$ (with $d\Omega^2 \equiv
d\theta^2 + \sin^2\theta \, d \varphi^2$ the metric of the unit sphere).
In these coordinates,
the metric form is, in general, non-diagonal. When, in addition,
the 3-surfaces defined by $t=constant$ are flat, the curvature
coordinate system is called a PG coordinate system.
From now on, it will be understood
that the title of the sections and the
results of this work always concern space-times with spherical
symmetry. The agreement for space-time signature is $(- + + + )$.

\section{Some geometrical relations} \label{sec2}

In this section we present the geometric background needed
in the following sections: expressions for the Ricci tensor
and the extrinsic curvature of a spherically symmetric synchronization, as
well as, the mean curvature vector and the Gauss identity for the
2-spheres of a SSST. Of course, this material is not new and it may be bypassed
or used as a glossary of formulae, which are conveniently referred
throughout  the text of the remaining sections. For an account on 2+2 warped space-times properties allowing to intrinsically characterize SSST see \cite{FeSa1,FeSa2}.

Let $(V_4, g)$ be a SSST, and let us consider a canonical coordinate system
$\{T, R, \theta, \varphi\}$ of $V_4$, which is adapted to the symmetries of the metric $g$.
Then we may express the metric in the following general form \cite{Synge,Takeno,PlebKra}
\begin{equation}\label{ds22}
g = A\, dT \otimes dT  + B \, dR \otimes dR + C (d T \otimes d R + d R \otimes d T) +  D \, \sigma,
\end{equation}
with $A(T,R), B(T,R), C(T,R)$ verifying the condition $\delta
\equiv AB-C^2 <0$, $D(T,R) \neq 0$, and $\sigma \equiv d\theta \otimes d \theta + \sin^2 \, \theta
d\varphi \otimes d \varphi$ being the metric on the unit $2$-sphere.

\subsection{Ricci tensor} \label{sec2A}
The induced metric
$\gamma$ on the $3$-surfaces $T=constant$, is written as
\begin{equation}\label{gamma}
\gamma = B dR \otimes d R + D  \sigma,
\end{equation}
with $B \neq 0$. The Ricci tensor of $\gamma$,
${\mathcal Ric}(\gamma)$, is given by
\begin{equation}\label{Ric}
{\mathcal Ric}(\gamma) = \left(\frac{B}{2} {\mathcal R} -
\frac{B}{D} F \right) dR \otimes d R +  \left(\frac{D}{4} {\mathcal
R} + \frac{F}{2} \right) \sigma
\end{equation}
where
\begin{equation}\label{F}
F= 1 - \frac{(\partial_R D)^2}{4 B D}
\end{equation}
and ${\mathcal R}\equiv {\mathcal R}(\gamma)$, the scalar
curvature of $\gamma$, results
\begin{equation}\label{Rdd}
\begin{array}{lcl}
{\mathcal R} & = & {\displaystyle\frac{2}{D}\left( 1 +
\frac{\partial_RB\partial_RD}{2B^2} + \frac{(\partial_RD)^2}{4BD} -
\frac{\partial_R^2D}{B} \right)} \\ \\
& = & \left\{\begin{array}{ll} {\displaystyle \frac{2F}{D} +
\frac{4\partial_RF}{\partial_RD}} & \quad {\rm if } \quad
\partial_RD\neq 0,
\\ [0.4cm] {\displaystyle\frac{2}{D}} & \quad {\rm if } \quad \partial_R D=0.\end{array}\right.
\end{array}
\end{equation}
Notice that the 3-surfaces $T=constant$ are conformally flat%
\footnote{Notice that the Cotton tensor of $\gamma$,
${\cal C}_{ijk}(\gamma) \equiv D_i Q_{jk}-D_j Q_{ik}$,
($D_i$ is the covariant derivative with the Levi-Civita
connection of $\gamma$, and $Q_{ij}\equiv {\mathcal R}_{ij}-\frac{\mathcal
R}{4}\gamma_{ij}$) identically vanishes, ${\mathcal C}_{ijk}(\gamma)=0$.
This is the result that one could expect ought to the algebraic properties of
the Cotton tensor and the assumed spherical symmetry. This means that
$\gamma$ is a conformally flat metric. Then, $\gamma$ may always be written
in isotropic form by making a coordinate change on each 3-surface $T=constant$.
An interesting summary on the Cotton tensor properties is given in \cite{GaHeHeMa}.
\label{fn:Cotton}}
but, in general, they are not flat.
From Eqs. (\ref{Ric}), (\ref{F}) and (\ref{Rdd}), we see that
$\gamma$ is a flat metric if, and only if, $F=0$, that is
\begin{equation}\label{plana}
{\mathcal Ric}(\gamma)=0  \quad \Longleftrightarrow \quad 4 B D =
(\partial_R D)^2.
\end{equation}

\subsection{Extrinsic curvature} \label{sec2B}

The extrinsic curvature of the slicing $T=constant$ is defined as $K
= - \frac{1}{2} {\cal L}_{n} \gamma$ where ${\cal L}_{n}$ is the Lie
derivative along the unit normal vector $n$,
\begin{equation}\label{normal}
    n = \frac{1}{\alpha}\left(\frac{\partial}{\partial T} - \frac{C}{B}\frac{\partial}{\partial R}\right)
\end{equation}
with $\alpha^2 = {- \displaystyle \frac{\delta}{B}}$. From
(\ref{gamma}), we obtain
\begin{equation}\label{K}
 K = \Psi B \, d R \otimes d R + \Phi  D \, \sigma
\end{equation}
where
\begin{equation}
 \Psi = K_R^R=\frac{1}{2B\alpha}\, \left( 2\partial_R C - \frac{C}{B}\partial_R B -\partial_T B\right)
\end{equation}
\begin{equation}
 \Phi = K_\theta^\theta = K_\varphi^\varphi =
 \frac{1}{2D\alpha}\, \left( \frac{C}{B}\partial_R D - \partial_T D\right)
\end{equation}
are the eigenvalues of $K$. Developing the Lie derivative of $K$
with respect to the shift vector $\beta = \frac{C}{B}
\frac{\partial}{\partial R}$ we arrive to the expression
\begin{equation}\label{LieK}
{\cal L}_\beta K = \frac{B}{C}\, \partial_R\left[
\left(\frac{C}{B}\right)^2 \Psi B\right] \, d R \otimes d R +
\frac{C}{B}\, \partial_\theta(\Phi  D) \, \sigma,
\end{equation}
which will be needed in Sec. \ref{sec6} to split the Einstein
evolution equations with respect to a PG synchronization.

\subsection{Mean curvature vector} \label{sec2C}

The mean curvature vector $H$ of a $2$-sphere $S$ defined by
constant $T$ and $R$ is given by
\begin{equation}\label{H}
H = - \frac{1}{D\delta }\Big[(B \partial_T D - C \partial_R D)
\frac{\partial}{\partial T} + (A \partial_R D - C \partial_T D)
\frac{\partial}{\partial R}\Big].
\end{equation}
This expression directly follows by taking the trace (with respect
to the induced metric $D \sigma$) of the extrinsic curvature tensor
${\cal K}$ of each $S$, which is defined by
\begin{equation}\label{K-tensor}
{\cal K}(e_a, e_b) \equiv - \left(\nabla_{e_a} e_b \right)^\perp
=- \left(\Gamma_{ab}^T \ \frac{\partial}{\partial T}
+\Gamma_{ab}^R \ \frac{\partial}{\partial R}\right)
\end{equation}
where the minus sign is taken as a matter of convention. $\nabla$
is the Levi-Civita connection of $g$, $\{e_i \equiv
\frac{\partial}{\partial x^i}\}$ is a coordinate basis of $S$,
$i=a,b=\theta,\varphi$, and the symbol $^\perp$ stands for the
projection on $S^\perp$: the time-like 2-surface orthogonal to $S$.
For a detailed study of ${\cal K}$ with applications in relativity
see, for example, \cite{JJA,Seno,MM}. Then, it results
\begin{equation}\label{KdeS}
{\cal K} = \frac{1}{2} D \sigma \otimes H.
\end{equation}
The one form $\Gamma$ metrically equivalent to $H$,
$\Gamma_\alpha = g_{\alpha \beta} H^\beta$, is written as
\begin{equation}\label{Gamma}
\Gamma = - \frac{1}{D}(\partial_T D  dT + \partial_R D dR) = - d
\ln D,
\end{equation}
which can be also obtained from the general expressions presented in \cite{Seno2}.

\subsection{Gauss identity} \label{sec2D}

Given a space-like 2-surface $\Sigma$ of a space-time
$(V_4, g)$,  the Gauss identity provides a scalar relation
involving the background geometry and the intrinsic and extrinsic properties of $\Sigma$,
and it may be expressed as%
\footnote{The Gauss identity is usually given in terms of the
Riemann curvature tensor $R_{\alpha\beta\gamma\delta}$ (see, for
instance \cite{Seno,MM}) from which the expression (\ref{Gauss})
follows by taking into account the algebraic decomposition:
$$
R_{\alpha\beta\gamma\delta} = W_{\alpha\beta\gamma\delta} +
g_{\alpha[\gamma}R_{\delta]\beta}-g_{\beta[\gamma}R_{\delta]\alpha}-
\frac{R}{3}\, g_{\alpha[\gamma}g_{\delta]\beta}
$$
where the bracket stands for antisymmetrization of index pairs,
$T_{[\alpha\beta]}\equiv\frac{1}{2}(T_{\alpha\beta}-T_{\beta\alpha})$.
 \label{fn:curvature}}
\begin{equation} \label{Gauss}
R(h) = \frac{2}{3} R (g) + g(H, H) + 2 tr (K_l \times
K_k) + 2 Ric(l,k) - 2 W(l, k, l, k) \quad
\end{equation}
where $R(h)$ is the scalar curvature of the induced metric $h$ on
$\Sigma$, $R(g)$, $Ric$ and $W$ are, respectively, the scalar
curvature, the Ricci and the Weyl tensor of $g$, and $l$ and $k$
are two independent future pointing null vectors normal to $\Sigma$
satisfying $g(l,k)=-1$. Now, $H$ is the mean curvature of
$\Sigma$, that is $H = tr {\cal{K}}$, with ${\cal{K}}$ the extrinsic
curvature tensor of $\Sigma$ (as defined by the first equality of
(\ref{K-tensor}), taking $e_a$ and $e_b$ tangent to $\Sigma$);
$K_l$ and $K_k$ are, respectively, the second fundamental forms of
$\Sigma$ relative to $l$ and $k$, that is, $K_l (e_a, e_b) = g(l,
{\cal{K}}(e_a,e_b))$, and similarly for $K_k$. The trace is taken
with respect to $h$, that is, $tr (K_l \times K_k) = h^{ad} h^{bc}
(K_l)_{ab} (K_k)_{cd}$.

For $\Sigma=S$, a 2-sphere (orbit) in a SSST, the Gauss identity
may be written as
\begin{eqnarray} \label{GaussSS}
\rho & = & 2 \lambda + \frac{1}{3}(4 \mu - \mu_1 - \mu_2)
+ \frac{1}{2} H^2.
\end{eqnarray}
with $\rho = 2/D$ the scalar curvature of $D \sigma$, $\lambda$ the simple eigenvalue%
\footnote{One has
$$
\frac{1}{2}W_{\mu\nu \alpha\beta}(l^\alpha k^\beta - l^\beta k^\alpha) =
\lambda (l_\mu k_\nu - l_\nu k_\mu),
$$
and then,  $W(l, k, l, k) \equiv W_{\mu\nu \alpha\beta}l^\mu k^\nu l^\alpha k^\beta = - \lambda$.
 \label{fn:Weyl-autovalor}}
of $W$ (provided that $W\neq 0$),
and  $\{\mu, \mu_1, \mu_2\}$ stand for the Ricci eigenvalues.
More precisely, the tangent space to $S$ is the eigenplane associated with
the eigenvalue $\mu$. The eigenvalues $\mu_1$ and
$\mu_2$ may be real (simple or multiple) or complex, and have
associated the invariant 2-plane tangent to $S^\perp$. Here, we
have considered the algebraic classification of
Churchill-Pleba\'{n}ski for a symmetric 2-tensor in Lorentzian
geometry, and its peculiarities in spherical symmetry (see
\cite{PlebSta}, and also \cite{BCM} for an intrinsic approach). We
have taken into account that $R(g)= 2 \mu + \mu_1 +
\mu_2$, and $2 Ric (l, k) = - (\mu_1 + \mu_2)$. In
addition $H^2 \equiv g(H,H)= - 4 tr (K_l \times K_k)$ because, according
to (\ref{KdeS}), the second fundamental forms $K_l$ and $K_k$
are both proportional to $h=D \sigma$.

\section{Painlev\'e-Gullstrand slicings}
\label{sec3}

In this section we find the condition ensuring the existence
of a PG-synchronization in spherical symmetry, and
we discuss its invariant meaning in terms of the eigenvalues of the Weyl and Ricci tensors.
Also, using a radial curvature coordinate $r$, we provide a physical interpretation
of this condition in terms of the Misner-Sharp gravitational energy of a 2-sphere of radius $r$.
As commented below, this result shows an interesting interconnection  between the study and classification
of trapped surfaces (see \cite{Hayward-96}) and the existence of flat slicings.

\subsection{Existence condition} \label{sec3A}

Let us start from the general metric form (\ref{ds22}).
Exploring the gauge freedom to make coordinate transformations of the form $T = T(t,r)$, $R=R(t,r)$,
we look for a function $t(T, R)$ whose level hypersurfaces $t=constant$ are Euclidean,
i. e.,  the induced metric is positive and flat. Under such a transformation, the metric
(\ref{ds22}) is expressed as
\begin{equation}\label{ds23}
 ds^2= \xi^2 \, dt^2 +  \chi^2 dr^2 + 2 \xi \cdot \chi  \, d t \, d r +  {\cal D}(t,r)
 d\Omega^2,
\end{equation}
with   ${\cal D}(t,r) \equiv D(T(t,r), R(t,r))$,  and the vector
fields $\xi$ and $\chi$ defined by
\begin{equation} \label{XiChi}
\xi \equiv \dot{T} \frac{\partial}{\partial T} + \dot{R}
\frac{\partial}{\partial R}, \qquad \chi \equiv T'
\frac{\partial}{\partial T} + R' \frac{\partial}{\partial R}.
\end{equation}
Over-dot and prime stand for partial derivative with respect $t$
and $r$, respectively, and $J \equiv \dot{T} R' - T' \dot{R} \neq
0$ assures coordinate regularity.

The scalar products $\xi^2 \equiv g(\xi,\xi)$, $\chi^2 \equiv
g(\chi,\chi)$ and $\xi \cdot \chi \equiv g(\xi,\chi)$  can be written as
\begin{equation}\label{xi2-c}
\delta (dD)^2 \, \dot{T}^2 + 2 \dot{{\cal D}} \, Z  \, \dot{T}  +
B \dot{{\cal D}}^2 = (\partial_R D)^2 \, \xi^2,
\end{equation}
\begin{equation}\label{chi2-c}
\delta (dD)^2 \, T'^2 + 2 {\cal D}' \, Z  \, T'  + B {\cal D}'^2 =
(\partial_R D)^2 \, \chi^2,
\end{equation}
\begin{equation}\label{xichi-c}
(\partial_R D)^2 \, \xi \cdot \chi =  \delta (dD)^2 \, \dot{T} \,
T' + Z \, ({\cal D}' \dot{T} + \dot{{\cal D}} T') + B \dot{{\cal
D}} {\cal D}' ,
\end{equation}
where we have used the relations $\dot{{\cal D}} = \dot{T} \partial_T D +
\dot{R} \partial_R D$ and ${\cal D}' = T'
\partial_T D + R' \partial_R D $ to substitute $\dot{R}$ and $R'$ in terms of $\dot{T}$ and $T'$.
We have denoted $Z \equiv C \partial_R D - B \partial_T D$. Then, due to
the Lorentzian character of the metric ($\delta <0$),
Eqs. (\ref{xi2-c}) and (\ref{chi2-c}) lead to real values for
$\dot{T}$ and $T'$ if, and only if,  the inequalities
\begin{equation}\label{des-dos}
\xi^2 (d {\cal D})^2 \leq {\dot{\cal D}}^2,
\quad \chi^2 (d {\cal D})^2 \leq {\cal D}'^2
\end{equation}
are satisfied.  Now, looking for a flat synchronization,  we have that the
induced metric on the 3-surfaces $t=constant$ is flat if, and only
if,
\begin{equation}\label{3flat}
4 \, {\cal D} \,  \chi^2 = {\cal D}'^2 \,
\end{equation}
according to Eq. (\ref{plana}).
Consequently, in the case of a flat synchronization, the second inequality in
(\ref{des-dos}) is equivalent to
\begin{equation}\label{des-chi-flat}
(d \sqrt {\cal D})^2 \leq 1.
\end{equation}
So, under the assumed spherical symmetry, Eq. (\ref{des-chi-flat})
provides the necessary and sufficient condition to be fulfilled for the existence
of a flat slicing. The first inequality in (\ref{des-dos}) guarantees
that the slices are space-like, that is, that the slicing is a PG
synchronization.

\subsection{Geometric interpretation} \label{sec3B}

In terms of the scalar curvature $\rho=2/{\cal D}$ of the metric ${\cal D}d\Omega^2$,
the above condition (\ref{des-chi-flat}) may be expressed as follows
\begin{equation}\label{invariant}
(d\rho)^2\leq 2\rho^3,
\end{equation}
which involves the sole invariant $\rho$. On the other hand,
according to Eq. (\ref{Gamma}), $H^2=\Gamma^2= (d \ln {\cal D})^2$,  and then Eq.
(\ref{des-chi-flat}) gives an upper bound for the norm of the mean
extrinsic curvature $H$ of the group orbits,
\begin{equation}\label{Hquadrat}
H^2 \leq \frac{4}{{\cal D}} = 2 \rho.
\end{equation}
Moreover, from the Gauss relation (\ref{GaussSS}), we arrive to the following result.

\begin{proposition}
In a  spherically symmetric space-time the following conditions are equivalent.
\begin{itemize}
\item[(i)] There exists a Painlev\'{e}-Gullstrand synchronization.
\item[(ii)] $(d\rho)^2\leq 2\rho^3$, where $\rho$ is the scalar curvature of the 2-spheres.
\item[(iii)] $H^2 \leq 2 \rho$, where $H$ is the mean curvature vector of the 2-spheres.
\item[(iv)]
$\mu_1 + \mu_2 - 4 \mu \leq 6 \lambda$,
where $\mu_1, \mu_2$, and $\mu $ (double) are the Ricci eigenvalues,
and $\lambda$ is the simple eigenvalue of the
Weyl tensor, or $\lambda = 0$ when the space-time is conformally flat.
\end{itemize}
\end{proposition}
Notice that this is a geometric result, which will be physically interpreted in the next subsection. Taking
into account the Einstein equations, in the above item (iv)
the Ricci eigenvalues, $\{\mu_1, \mu_2, \mu\}$  may by substituted by
the corresponding energy tensor eigenvalues, $\{e_1, e_2, e\}$, giving
\begin{equation}\label{enT}
e_1 + e_2 - e \leq 3 \lambda.
\end{equation}

\subsection{Physical interpretation} \label{sec3C}

By definition, see \cite{Takeno,Synge},   $r$ is a coordinate of
curvature  for the spherically symmetric metric form (\ref{ds23})
if ${\cal D}(t,r)= r^2$, so that  ${\cal D}' = 2r$ and $\dot{{\cal
D}} = 0$. Then, (\ref{des-chi-flat}) says that $(dr)^2 \leq 1$,
and taking into account that the Misner-Sharp gravitational energy
$E$ of a 2-sphere of radius $r$ is expressed as (see
\cite{MiSha,Hayward-96})
\begin{equation}\label{MS-energy}
E = \frac{r}{2}\Big(1- (dr)^2\Big),
\end{equation}
we arrive to the following result.

\begin{proposition}
Any spherically symmetric space-time admits
a Painlev\'e-Gullstrand  synchronization in the region
where the Misner-Sharp gravitational energy is non-negative, $E
\geq 0$.
\end{proposition}

The Misner-Sharp energy has been painstakingly analyzed in
\cite{Hayward-96}, providing useful criteria to study
trapped surfaces. The main novelty here has been to relate this concept
and the existence of PG synchronizations.%
\footnote{Marc Mars inspired us in the obtaining of this relation.}
\label{fn:MarcM}

Moreover, the flatness condition (\ref{3flat}) implies that $\chi^2 = 1$,
and the metric (\ref{ds23}) is written as
\begin{equation}\label{ds24}
 ds^2= \xi^2\, dt^2 + 2 \xi \cdot \chi  \, d t \, d r + d r^2 + r^2
 d\Omega^2.
\end{equation}
Then, accordingly to (\ref{xi2-c}) and (\ref{xichi-c}), the following
relations must occur
\begin{equation}\label{AxiB}
\begin{array}{l}
{\cal A} (t, r) \equiv \xi^2 = J^2 \, \delta  (dr)^2 \\ \\
{\cal B} (t, r) \equiv \xi \cdot \chi = \varepsilon J \sqrt{\delta
[(dr)^2 -1]}
\end{array}
\end{equation}
where $\varepsilon=\pm 1$. So, the real function ${\cal B}$ exists in
the region where $(dr)^2 \leq 1$, and we have the following
result.

\begin{proposition}
Let $r$ be the radius of curvature of the
orbits ($2$-spheres) of the isometry group of a spherically
symmetric space-time with metric $g$. In the region defined by the condition
\begin{equation}\label{dr}
(dr)^2 \equiv g^{\mu\nu} \partial_\mu r \partial_\nu r \leq 1,
\end{equation}
the Misner-Sharp energy is not negative and a curvature coordinate system
$\{t, r, \theta, \varphi\}$ exists in
which the metric line element may be written as
\begin{equation}\label{ds21}
d s^2= {\cal A} (t, r)\, dt^2 + 2 {\cal B}(t, r) \, d t \, d r + d
r^2 + r^2 d\Omega^2,
\end{equation}
where $d\Omega^2 = d\theta^2 + \sin^2 \theta \, d\varphi^2$.
\end{proposition}

The Lorentzian character of the metric impose that the functions
${\cal A}$ and ${\cal B}$ must satisfy the sole restriction ${\cal
A}< {\cal B}^2$, which is implied by (\ref{AxiB}).

The Misner-Sharp energy is a geometric invariant that may be
physically interpreted as an effective gravitational energy whose origin
is the interaction between the energetic content
and its associated field (see \cite{Hayward-96}).
Given that the intrinsic and extrinsic scalar curvatures of the 2-spheres are
$\rho = 2/r^2$ and $H^2 = 4(d \ln r)^2$, and according with (\ref{MS-energy})
one has the invariant expression
\begin{equation}\label{MS-energy2}
E = \frac{1}{\sqrt{2\rho}}\Big(1- \frac{1}{2\rho}H^2\Big).
\end{equation}
Finally, notice that Eq. (\ref{Hquadrat}) does not constraint the
causal character of the mean curvature vector $H$, which might be
time-like, light-like or space-like. This is a remarkable
property, because a 2-sphere is said to be trapped, marginal or
untrapped if $H$ is, respectively, time-like, light-like or
space-like (see e. g. \cite{Seno,MM,Hayward-96,Hayward-10}).

\section{Energy and momenta densities of a Painlev\'e-Gullstrand slice}
\label{sec4}

In this section we establish the following result.

\begin{proposition}
In any spherically symmetric space-time, the
Weinberg energy and momenta densities vanish for every
Painlev\'e-Gullstrand synchronization.
\end{proposition}

Of course, to find a coordinate system in which the
Weinberg densities vanish is not a surprising property,
due to the non-tensorial character of them. However, the novelty here is to show that,
for every SSST,  such a vanishing property occurs
in PG coordinates.

In order to proof the above result, let us consider the metric (\ref{ds21}) written in a
quasi-Minkowskian form, that is $g=\eta +h$ with $\eta$ the
Minkowski metric, $h_{00}=1+{\cal A}$, $h_{0i}={\cal B}x_i/r$, and
$h_{ij}=0$.

We start from the expression of the Weinberg pseudo-tensor
\cite{Wein},
\begin{equation*}
2Q^{i0\lambda} = \frac{\partial h_\mu^\mu}{\partial
x_0}\eta^{i\lambda} - \frac{\partial h_\mu^\mu}{\partial
x_i}\eta^{0\lambda} - \frac{\partial h^{\mu 0}}{\partial
x^\mu}\eta^{i\lambda} + \frac{\partial h^{\mu i}}{\partial
x^\mu}\eta^{0\lambda} + \frac{\partial h^{0\lambda}}{\partial x_i} - \frac{\partial
h^{i\lambda}}{\partial x_0} ,
\end{equation*}
where Latin and Greek indexes go from 1 to 3 and from 0 to 3,
respectively, and all indexes are raised and lowered with the flat
metric $\eta$. In this case, it results $Q^{i00}=0$ (according with \cite{MirAb}) and
\begin{equation}\label{q}
2Q^{i0j}=\left(\frac{\cal B}{r}+{\cal B}^\prime\right)\delta_{ij} +
\left(\frac{\cal B}{r}-{\cal
B}^\prime\right)\frac{x_i}{r}\frac{x_j}{r}
\end{equation}

The derivative of this expression leads to
\begin{eqnarray*}
2 \ \frac{\partial Q^{i0j}}{\partial x^k} & = &\left( \frac{{\cal
B}^\prime}{r} -\frac{\cal B}{r^2} + {\cal B}^{\prime\prime}
\right)\delta_{ij} \ \frac{x_k}{r} +
\left(\frac{\cal B}{r^2}-\frac{{\cal B}^\prime}{r}\right)\left(\delta_{ik} \ \frac{x_j}{r}
+\delta_{jk} \ \frac{x_i}{r}\right) \\ & &
+\left(3\frac{{\cal B}^\prime}{r}-3\frac{\cal B}{r^2}-{\cal B}^{\prime\prime}\right)
\frac{x_i}{r} \ \frac{x_j}{r} \ \frac{x_k}{r}.
\end{eqnarray*}
and by contraction of the indexes, directly follows that $\frac{\partial
Q^{i0j}}{\partial x^i} =0$. Then, the four-momentum density
vanishes,
\begin{equation}
 \tau^{0\lambda}\equiv -\frac{1}{8\pi G}\frac{\partial
 Q^{i0\lambda}}{\partial x^i} =0
\end{equation}
and hence, the angular momentum densities ${\rm
j}^{i\lambda}=x^i\tau^{0\lambda}-x^\lambda\tau^{0i}$ also vanish, according
with the announced conclusion.

For the special case of the Schwarzschild geometry, the vanishing of the energy density
may be intuitively understood invoking the Einstein equivalence principle.
Taking $\epsilon = 1$ in the extended form (\ref{PG-form}) of the Schwarzschild metric, $t$ represents
the proper time of a radial geodesic observer which initially stays, in $r = \infty$,
at rest with respect to a static observer. Locally, such an observer
does not feel any gravitational effect.

\section{Painlev\'e-Gullstrand slicings and semi-metric connections}
\label{sec5}

In the eighties, Bel proposed an extended Newtonian theory of
gravitation based on a semi-metric connection associated with an
observer congruence and a flat spatial 3-metric \cite{Bel}. In a
space-time, with metric $g_{\mu\nu}$, which admits a spatially flat
slicing given by the coordinate hypersurfaces $x^0=constant$, the
connection coefficients of the aforementioned semi-metric connection
are written as \cite{Bel},
\begin{eqnarray}
\Lambda^k & = & -\Gamma_{00}^{k} = \frac{1}{2}\delta^{ki}(\partial_i g_{00} - 2\partial_0 g_{0i}) \\
\Omega_j^k & = & -2\Gamma_{0j}^k = \delta^{ki}(\partial_i g_{0j} -
2\partial_j g_{0i}).
\end{eqnarray}

Consequently, a SSST metric admits a Newtonian interpretation when
it is written in PG coordinates and it is considered in the above
context. In fact, taking into account the expression (\ref{ds21}) of
the metric, we have $g_{00}={\cal A}$, $g_{0i}={\cal B} x_i / r$,
and then
$$
\partial_i g_{0j} =\frac{\cal B}{r}\delta_{ij} +\left({\cal B}^\prime -\frac{\cal B}{r}\right)\frac{x_i}{r} \
\frac{x_j}{r}= \partial_j g_{0i}.
$$
Then, the connection coefficients result
\begin{eqnarray}
\Lambda^k & = &  \frac{1}{2} \Big({\cal A}^\prime - 2\dot{\cal B}\Big) \ \frac{x^k}{r} \\
\Omega_j^k & = & 0,
\end{eqnarray}
which means that, in the region of a SSST where a PG synchronization exist, the gravitational field
may be interpreted as an inertial field of radial accelerations and vanishing rotation.

In particular, for the case of the Schwarzschild metric, we have
${\cal A}= -\left( 1 -\frac{2m}{r}\right)$, ${\cal B}= \epsilon
\sqrt{\frac{2m}{r}}$, and then the vector component of the
connection reduces to
\begin{equation}\label{lliure}
\vec{\Lambda} = - \frac{m}{r^2} \vec{e_r},
\end{equation}
where $\vec{e_r}$ is the unit vector in the radial direction. The
above expression (\ref{lliure}) gives the acceleration of a unit
mass particle radially falling in the Newtonian field of a mass $m$.
Similar Newtonian interpretations have been considered from a
different point of view (see, for example,
\cite{MarPoisson,HamLis,Visser2}).

\section{Painlev\'e-Gullstrand slicings and Einstein equations}
\label{sec6}

In General Relativity, when dealing with the evolution (or
 $3+1)$ formalism (see \cite{Tesi-Tolo,Tolo-Bona}, and \cite{Gourgoulhon} for a recent
review) one introduces a  vorticity free observer $n$, $n^2=-1$, and
Einstein equations are decomposed in the following set of constraint
equations ($\kappa$ is the Einstein constant),
\begin{equation}\label{c1}
{\mathcal R}(\gamma) + ({\rm tr} K)^2 - {\rm tr} K^2 = 2\, \kappa \,
\tau\
\end{equation}
\begin{equation}\label{c2}
 \nabla \cdot (K - {\rm tr K} \, \gamma) = \kappa \, q
\end{equation}
and this other system of evolution equations
\begin{equation}\label{e1}
\partial_t \gamma = - 2 \alpha K + {\cal L}_{\beta} \gamma
\end{equation}
\begin{equation}\label{e2}
\partial_t K   = - \nabla \nabla \alpha - \kappa \alpha[\Pi +
\frac{1}{2}(\tau - p) \gamma]
+ \alpha [{\mathcal Ric}(\gamma) + {\rm tr} K \; K - 2 K^2] + {\cal
   L}_{\beta}K \,. \quad
 \end{equation}
Here, $\gamma$ and $K$ are, respectively, the metric and the
extrinsic curvature of the space-like slices whose normal vector
is $n$; $\nabla$ is the Levi-Civita connection of $\gamma$, and
the Ricci tensor and scalar curvature of $\gamma$ are denoted by ${\mathcal
Ric}(\gamma)$ and ${\mathcal R}(\gamma)$, respectively; the trace
operator associated with $\gamma$ is denoted by ${\rm tr}$, so that,
$(\nabla \cdot K)_a \equiv ({\rm tr} \nabla K)_a \equiv
\gamma^{ij}\nabla_i K_{ja}$ is the divergence of $K$ with
respect to $\gamma$. In the usual evolution formalism notation,
$n$ is written as $n=\alpha^{-1}(\frac{\partial}{\partial t} -
\beta)$,  where $\alpha$ is the lapse function and $\beta$ is the
shift vector.

The energy content ${\cal T} \equiv \{\tau, q, p, \Pi\}$ has been
decomposed relatively to $n$, that is
\begin{equation}
{\cal T} = \tau n \otimes n + n \otimes q + q \otimes n + \Pi + p
\gamma ,
\end{equation}
with $\tau \equiv {\cal T}(n,n)$, $q \equiv - \bot {\cal T}(n, \cdot)$, $p$
and $\Pi$ being the energy density,  the energy flux,
the mean pressure and the traceless anisotropic pressure as measured by $n$, respectively; $\bot $
is the projector on the 3-space orthogonal to $n$
associated with the 3-metric $\gamma \equiv g + n \otimes n$.

\subsection{Spherical symmetry}
\label{6A}

In the case of a SSST, using the expression (\ref{K}) of the
extrinsic curvature, the constraint equations (\ref{c1}) and
(\ref{c2}) are equivalent to
\begin{eqnarray}\label{c12}
\qquad \qquad \;\Phi(\Phi+2\Psi) & = & \kappa\tau - {\displaystyle\frac{{\mathcal R}}{2}}  \\
2\partial_R\Phi +{\displaystyle\frac{\partial_R D}{D}}(\Phi-\Psi) & = & -\kappa q_R
\label{c22}
\end{eqnarray}
where $q_R$ is now the radial component of the energy flux. For the
evolution equation (\ref{e2}), taking into account the expression
(\ref{LieK}), we have
$\Pi_{\varphi\varphi}=\Pi_{\theta\theta}\sin^2\theta$ and
\begin{equation}\label{e21}
\begin{array}{lcl}
\partial_T(\Psi B) & = & -\sqrt{B} \partial_R\left(
{\displaystyle\frac{\partial_R\alpha}{\sqrt{B}}}\right) -
   \kappa\alpha \Big( \Pi_{RR} +
{\displaystyle\frac{1}{2}}(\tau -p)B \Big) \\ [4mm]
  & & + \alpha\Big(
{\displaystyle\frac{B}{2}}{\mathcal R}
   - {\displaystyle\frac{B}{D}}F + B\Psi(2\Phi-\Psi)\Big) + {\displaystyle\frac{B}{C}}\partial_R\left[\left(
{\displaystyle\frac{C}{B}}\right)^2B\Psi\right]
   \end{array}
\end{equation}
\begin{equation}\label{e22}
\begin{array}{lcl}
\partial_T(\Phi D) & = & {\displaystyle-\frac{\partial_R D}{2B}} \partial_R\alpha -
   \kappa\alpha \Big( \Pi_{\theta\theta} +\frac{1}{2}(\tau -p)D \Big) \\ [4mm]
   & & + \displaystyle{\alpha\Big(\frac{D}{4}{\mathcal R}+
   \frac{F}{2} + \Phi\Psi D\Big) + \frac{C}{B}\partial_R (D\Phi)}.
\end{array}
\end{equation}
For the metric (\ref{ds22}), Eqs. (\ref{c12}), (\ref{c22}),
(\ref{e21}) and (\ref{e22}) are the $3+1$ splitting of the Einstein
equations with respect to a vorticity free observer.  The proper
space of such an observer is Euclidean if, and only if,  $F=0$,
and then ${\cal R} = 0$. The integration of these equations for simple
energetic contents (for instance, a dust model) should provide the corresponding metric form in
PG coordinates. In the next section, the
vacuum case is considered: the extended form of Schwarzschild solution
in PG coordinates is obtained from the sole consideration
of the field equations.

\subsection{Schwarzschild vacuum solution}
\label{6B}

The extended Painlev\'e-Gullstrand-Lema\^{\i}tre
metric form of the Schwarzschild solution
may be obtained assuming spherical symmetry and the existence of a
flat synchronization, ${\mathcal Ric}(\gamma)=0$, and then, solving
the vacuum Einstein equations in a coordinate system adapted to such a
synchronization. So, let us take $\tau = p = \Pi_{RR} =
\Pi_{\theta\theta}= F = {\cal R} =0$. Then, for the metric expression
(\ref{ds21}), the lapse function is given by $\alpha^2={\cal
B}^2-{\cal A}$ and the constraint equations (\ref{c12}) and
(\ref{c22}) result
\begin{eqnarray}
\Phi(\Phi+2\Psi) & = & 0 \label{c13}\\
r\Phi' +\Phi-\Psi & = & 0 \label{c23}
\end{eqnarray}
with
\begin{equation}\label{PhiPsi}
 \Phi = \frac{1}{\alpha}\frac{\mathcal B}{r}, \quad \quad
 \Psi = \frac{{\mathcal B}'}{\alpha} \, .
\end{equation}

When $\Phi=0$, taking into account also the evolution equations, we
recover the Minkowski space-time. In the generic case,  $\Phi = -2
\Psi \neq 0$, Eq. (\ref{PhiPsi}) leads to
\begin{equation}\label{Befe}
{\mathcal B} = f(t) r^{-1/2}
\end{equation}
with $f(t)$ an arbitrary function. Substituting Eq. (\ref{Befe})
in the momentum constraint (\ref{c23}), it reduces to $\alpha' = 0$.
Consequently, the lapse is a function of the sole variable $t$,
$\alpha(t)$, and we can take $\alpha =1$ by re-scaling the coordinate $t$.
Then, we have
\begin{equation}\label{PhiPsi2}
\Phi = f(t) r^{-3/2}= -2 \Psi .
\end{equation}
Next, the evolution equations (\ref{e21}) and (\ref{e22}) are written as
$$
\dot{\Psi}=\Psi(2\Phi-\Psi)+\frac{1}{{\cal B}}({\cal B}^2\Psi)^\prime
$$
$$
\dot{\Phi}=\Psi\Phi+\frac{\cal B}{r^2}(r^2\Phi)' .
$$
Given that $\Phi=-2\Psi$, these last equations are equivalent to
\begin{equation}\label{first}
\dot{\Psi}=\Psi^2+\frac{\cal B}{r^2}(r^2\Psi)'
\end{equation}
\begin{equation}\label{second}
3\Psi+\frac{\cal B}{r}-{\cal B}'=0 .
\end{equation}
By using the expressions (\ref{Befe}) and (\ref{PhiPsi2}), the equation
(\ref{first}) leads to $f(t)=constant$ and the equation
(\ref{second}) is identically satisfied. Finally, by taking $f= \epsilon
\sqrt{2m}$, we obtain
\begin{equation}\label{PG-form}
 ds^2=- \left(1-\frac{2m}{r}\right)\, dt^2 + 2 \epsilon \sqrt{\frac
{2m}{r}} \, d t \, d r + d r^2 + r^2 d\Omega^2,
\end{equation}
which is the extended form of the Schwarzschild solution obtained
by Painlev\'e, Gullstrand and Lema\^{\i}tre
\cite{Painleve,Gullstrand,Elmestre}. The positive parameter $m$ is the
Schwarzschild energy. The sign $\epsilon$ provides two coordinate branches for the solution,
the Kruskal-Szekeres black and white hole regions
being described by the above metric with $\epsilon = 1$  and $\epsilon
=-1$, respectively, see \cite{Schutzhold}. Note that the $r$ coordinate can take any positive value,
$0<r<+\infty$. In fact, from (\ref{PG-form}), we have $(dr)^2=
g^{rr}= 1-\frac{2m}{r} < 1$, and the domain of a
Painlev\'e-Gullstrand chart extends for every value of $r \neq 0$.
Notice that (\ref{MS-energy}) implies that $E=m$, which provides the physical interpretation of the
parameter $m$ as an effective energy \cite{Hayward-96}. Moreover, writing
$dt = dt_S + \epsilon \sqrt{\frac{2m}{r}}(1-\frac{2m}{r})^{-1} dr$,
one recovers the usual Schwarzschild metric form
\begin{equation}\label{Sch-form}
 ds^2=- \left(1-\frac{2m}{r}\right) dt_S^2 + \left(\displaystyle{1- \frac
{2m}{r}}\right)^{-1} d r^2 + r^2 d\Omega^2
\end{equation}
where $t_S$ is the coordinate time of the static observer $(-\infty
< t_S < \infty)$ and the rank of the $r$ coordinate is restricted to
be $r > 2m$. According to the Jebsen-Birkhoff theorem (see \cite{Jebsen-Birkhoff} an references therein),
we recover the Schwarzschild metric as the sole spherically symmetric solution
of the vacuum Einstein equations.

Other derivations of the Schwarzschild solution providing  improvements of the original
proof of the Jebsen-Birkhoff theorem have been achieved by solving the field equations
in null coordinates (see \cite{Hayward-96, Lake-06} and reference therein).
From the conceptual point of view, any of these derivations make unnecessary to get
a coordinate transformation allowing to extend the domain of Schwarzschild
chart from the outer to the inner horizon regions.

\section{Discussion}
\label{sec7}

In this work we have analyzed the existence of flat
synchronizations in SSSTs. The condition (\ref{Hquadrat})
provides an upper bound for the norm of the mean extrinsic
curvature vector of the isometry group orbits which,
using the Gauss identity, may be expressed in terms
of curvature invariants. Moreover, the
associated flat slices have vanishing Weinberg energy and momentum densities.
We have seen that any spherically symmetric metric admits a Newtonian interpretation
in the context of the Bel extended Newtonian theory of gravitation. Our study offers a
new perspective about the meaning of the Painlev\'e-Gullstrand coordinates. This study
applies for any SSST in the region where these coordinates exist.
In this region, the gradient of the radial Painlev\'e-Gullstrand
coordinate $r$ may be space-like, light-like,  or time-like, according to
the condition $(dr)^2\leq 1$, which means that the Misner-Sharp
gravitational energy of a sphere of radius $r$ is non-negative.
This condition may be tested for any SSST, starting from the
general metric form (\ref{ds22}). For
instance, it occurs elsewhere in the Schwarzschild geometry, as
it has been pointed out at the end of Sec. \ref{sec6}.
Moreover, one has that $(dr)^2 \leq 1$ everywhere for any
Robertson-Walker metric with an energetic content
which satisfies the usual energy conditions. In fact,
if we put in (\ref{ds22}) $A=-1$, $B=a^2(t)/(1 + \frac{k}{4}
r^2)^2$ (with $k = 1, 0, -1$ the
universe curvature index), $C=0$ and $D=r^2 B$, we obtain that, in this case,
(\ref{Hquadrat}) is equivalent to $k + \dot{a}^2 \geq 0$, which
means that the proper energy density of the cosmological fluid is
non negative. Consequently, any Robertson-Walker space-time that satisfies this
energy condition admits a PG synchronization. This property is also
obtained directly from the inequality (\ref{enT}).
In this case,  $\tau=-e_1$ and $p=e_2 = e$ are, respectively,
the energy density and the pressure of the cosmological fluid,
and $\lambda =0$, because the Robertson-Walker metric is conformally flat.
We leave for a future work the obtaining of the Painlev\'e-Gullstrand form
of these Robertson-Walker cosmological models.

Finally, we have presented an improved proof of the Jebsen-Birkhoff theorem
by expressing and solving the vacuum Einstein Equations in PG-coordinates.
So, the extended Painlev\'e-Gullstrand-Lema{\^{\i}}tre
metric form of the Schwarzschild solution is directly obtained.

\ack
We appreciate valuable aids provided by the tandem Joan J. Ferrando--Juan A. S\'{a}ez
during the development of this paper, and the  orientational comment by Marc Mars during
the presentation of this subject at the Spanish Relativity Meeting ERE-2009.
This work has been supported by the
Spanish Ministerio de Ciencia e Innovaci\'on MICIN-FEDER project
FIS2009-07705.

%
\section*{References}

\end{document}